# AN INTEGRATED FRAMEWORK FOR DEVELOPING AND EVALUATING AN AUTOMATED LECTURE STYLE ASSESSMENT SYSTEM

Eleni Dimitriadou[0000-0001-9257-0999]   Andreas Lanitis[0000-0001-6841-8065]
Visual Media Computing Lab
Department of Multimedia and Graphic Arts
Cyprus University of Technology, Limassol, Cyprus.
CYENS Centre of Excellence, Nicosia, Cyprus
ela.dimitriadou@edu.cut.ac.cy, andreas.lanitis@cut.ac.cy.

## Abstract

The aim of the work presented in this paper is to develop and evaluate an integrated system that provides automated lecture style evaluation, allowing teachers to get instant feedback related to the goodness of their lecturing style. The proposed system aims to promote improvement of lecture quality, that could upgrade the overall student learning experience. The proposed application utilizes specific measurable biometric characteristics, such as facial expressions, body activity, speech rate and intonation, hand movement, and facial pose, extracted from a video showing the lecturer from the audience point of view. Measurable biometric features extracted during a lecture are combined to provide teachers with a score reflecting lecture style quality both at frame rate and by providing lecture quality metrics for the whole lecture. The acceptance of the proposed lecture style evaluation system was evaluated by chief education officers, teachers and students regarding the functionality, usefulness of the application, and possible improvements. The results indicate that participants found the application novel and useful in providing automated feedback regarding lecture quality. Furthermore, the performance evaluation of the proposed system was compared with the performance of humans in the task of lecture style evaluation. Results indicate that the proposed system not only achieves similar performance to human observers, but in some cases, it outperforms them.

**Keywords :** *Automated Lecture style assessment, Lecture style quality, Biometric features, Performance evaluation, Human performance.*

# 1 Introduction

Automated lecture quality assessment tools can provide objective and timely feedback on the quality of lecture delivery, assisting in that way educators to improve their lecture quality that will eventually lead to an improved learning experiences for students. In this study, we aim to develop an integrated application that provides in real time automated lecture delivery quality assessment, using measurable biometric characteristics extracted through video and audio recordings. The proposed application estimates lecture quality based on specific measurable biometric characteristics defined through a systematic multi-phased process that involves the definition of a set of lecture style quality indicators that reflect the goodness of a lecture presentation, as outlined in Figure 1. The biometric characteristics considered relate to facial expressions, body activity, speech rate and tone, hand movements, and facial pose, that can be extracted from video recordings of lectures either in real time or in a batch processing mode. Although, in theory the lecture quality could be estimated with deep networks that extract automatically features associated with good lecturing, the lack of suitable training data, requires the intermediate step of formulating a set of measurable biometric characteristics necessary for this task. The main steps in the process of defining suitable quality metrics include the step of requirements analysis where through a literature review and interviews with stakeholders, a set of biometric features associated with lecture delivery quality are defined, followed by the development of dedicated methods for extracting the features from a video stream. Extracted features are then used for estimating an overall lecture delivery quality score.

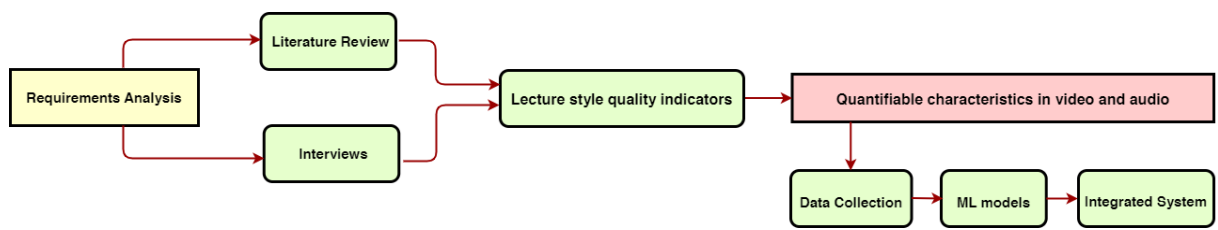

**Figure 1.** The systematic approach for developing a lecture quality assessment tool.

Although there are some existing tools that use automated methods to assess lecture quality [28], they have limitations in terms of the accuracy and range of attributes measured. Previous approaches for developing lecture evaluation systems include the work of Zhao and Tang [41] who describe the development of an automated classroom observation system to evaluate teacher performance. The proposed system uses video cameras and audio sensors to collect data from the classroom environment and machine learning algorithms to analyze the data and extract features such as teacher's speaking rate, student engagement, and teacher's body language. The extracted features are then used to generate a quantitative score of the teacher's performance.

Compared to previous work reported in the literature, in our approach we adopt a systematic approach for obtaining a comprehensive set of lecture quality indicators and then derive an integrated set of features associated with a good lecture rather than choosing features arbitrarily. Furthermore, the combination of quality scores from all features both at frame rate, and for the whole lecture, allow lecturers to get useful feedback regarding the lecture delivery quality both at real time, and as a report concerning the whole lecture. In relation to the biometric features used, when compared to previous efforts, in our work a more integrated set of features that includes facial pose, activity-related metrics, speech intonation, word density and hand movements, are considered. These measurable biometric characteristics of a good educator have been defined based on the current bibliographic material and the interviews of the persons directly involved in class delivery. The main difference with previous studies is that the metrics are defined based on the stakeholders, allowing the proposed application to better reflect the needs and requirements of the target user group. Using a combination of speech recognition and computer vision technologies, we aim to provide a more accurate and reliable assessment of the quality of teachers' lectures that captures both verbal and non-verbal cues. Furthermore, previous work reported in the literature does not include quantitative performance evaluation. In the work reported in this paper performance evaluation results of the proposed system are presented, and most importantly the results of a comparison of the performance of humans against the performance of the proposed automated lecture style evaluation system is also presented.

In a nutshell the main contributions of our work are:
- Utilizing human expertise for defining lecture style quality features.
- Use of an extended set of lecture style quality features.
- Performance evaluation and comparison with performance of humans.

In relation to our previous work in the area [15] which focuses on the developing an application for lecture quality assessment, in the present study we present the integrated lecture quality evaluation system, along with a novel performance evaluation methodology that allows the derivation of concrete conclusions related to the performance of the proposed system. Furthermore, a comparison between the performance of humans and machines in the task of lecture quality evaluation is presented.

In the remainder of the paper, we present a literature review on the topic of educator performance assessment (Section 2) and in Section 3 we outline the process of defining lecture quality indicators. In Section 4, the extraction of lecture quality indicators from video, and the estimation of the overall lecture quality score is discussed, followed by the presentation of the performance evaluation of our proposed system (Section 5). Section 6 includes concluding comments followed by a discussion and plans for future work.

## 2 Literature Review

Performance assessment of a lecturer is highly important as a means of safeguarding the quality of lectures delivered to students. Traditional lecturer performance assessment, based on lecture observation or student feedback, has been a quite complex and time-consuming process that does not provide instantaneous and seamless feedback. In this literature review a brief description of traditional and automated teacher performance assessment methods are presented.

## 2.1 Educator performance assessment

Educator performance assessment can be defined as a framework utilized in a classroom to assess and evaluate the performance of educators [23]. Traditional assessment methods are usually based on the observation of teachers by experts during course time something that can be expensive, time consuming, not accurate and usually the feedback provided is infrequent and is related to the performance and not on how teachers can enhance their techniques [3]. Furthermore, the use of student feedback to evaluate the teacher's performance is not a reliable method, as the data is not collected in real-time and responses can be manipulated [38]. To overcome this crucial impediment in teacher development, new technologies can be used to produce high quality and meaningful automatic lecture quality feedback for educators. Within this context lecture quality assessment refers to the evaluation of the lecturing style, rather than an overall teaching performance of a teacher that includes other aspects, such as teaching methodology, and subject knowledge.

In the context of assessing teaching quality in schools, colleges and universities, a variety of evaluation categories are often considered to provide a comprehensive view of the educational experience (see Table 1). These categories may include the subject knowledge [30], knowledge of learners [18], teaching methodology [12], curriculum knowledge [7], general pedagogical knowledge [26] , knowledge of contexts [13], self-knowledge [31], lecture style and audience interaction [33]. In essence, the effectiveness of a teacher is determined by the interconnectedness of these knowledge categories, with the ability to integrate and apply them holistically to facilitate optimal student learning and development.

Table 1. The evaluation categories for a teacher.

| Teachers' evaluation categories | Description |
|---|---|
| Subject Knowledge | A teacher's in-depth understanding of the subject matter they are teaching, including its content, principles, and methodologies. |
| Knowledge of Learners | Teachers should possess knowledge about students' biological, social, psychological, and cognitive development. |
| Teaching Methodology | A skilled teacher is well-versed in various teaching methods, strategies, and approaches. |
| Curriculum Knowledge | Teachers need to be familiar with the curriculum, textbooks, and educational policies relevant to their subject and grade level. |
| General Pedagogical Knowledge | Encompasses broader principles of classroom management, organization, and learning theory that extend beyond subject-specific content. |
| Knowledge of Contexts | Effective teachers are aware of the broader educational, social, and cultural contexts in which they work. |

| Self-Knowledge | Teachers should have a reflective understanding of their role, responsibilities, values, and professional development. |
|---|---|
| Lecture Style and Audience Interaction | A skilled teacher can deliver lectures clearly and engagingly, employing various teaching methods to maintain student interest and understanding. |

## 2.2 Automated teacher performance evaluation

Usually technologies used for educational performance assessment are equipped and updated with artificial intelligence techniques in order to provide accurate and objective results. Srivastava et al. [35] proposed an evaluation framework for assessing and improving teaching methods through the use of Internet of Things (IoT) systems, cloud computing and Machine Learning models. Moreover, IoT systems supported by artificial intelligence models such as fog-computing has stimulated new studies in the field [9]. Fog computing is a cloud-additional environment that can deliver information instantly [10]. The combination of IoT, Fog and Cloud computing improves the capabilities of many systems used in academic institutions. Furthermore, IoT-fog-cloud computing offers an important application for providing an intelligent educational experience and assessing students and teacher instantly.

Bhatia et al. [8] use IoT systems in classes to collect information regarding the performance of students and educators in order to identify their progress. Student data collected relates to different student activities performed in the school environment (e.g. academic performance, attendance, team-work) while teachers' data relates to the evaluation of their performance (e.g. quality of content, student satisfaction, number of assignments). Utilizing the Bayesian modelling approach, the collected information is assessed through a fog-cloud computing device with the aim to determine a quantifiable measure of performance. Furthermore, this progress measure is computed over time for both students' and educators' performance. The results of this method are viewed through the experiments conducted using four datasets and prove the efficiency of the method.

Yang et al. [40] proposed a teaching assessment model to address the complexities involved in evaluating teaching quality. In the proposed teaching model, the aim of using back propagation is to quantitatively measure the concept of teaching evaluation index, using specific data as input and teaching effectiveness as the output. The features used for teaching quality evaluation are related with guiding ideology, teachers, teaching conditions, teaching construction and reform, construction of study style, teaching management, teaching effectiveness, classroom teaching effectiveness, and others. The results reported were satisfactory, indicating that the model has a broad applicability in assessing teaching effectiveness. While the research presents positive outcomes, several aspects of their approach warrant further scrutiny. One notable limitation is the absence of comprehensive performance metrics to objectively evaluate the model's effectiveness. Although the authors assert that the results were satisfactory and the neural network outputs closely align with actual target values, the lack of specific metrics such as accuracy, precision, recall, or mean squared error undermines the quantitative assessment of the model's performance. The study would benefit from a comparative analysis with other established teaching quality assessment methods or benchmarks to ascertain the model's superiority and unique contributions.

Jensen, et al. [23] devised a method for teachers to effortlessly audiotape the conversations and lectures in a classroom. They also utilized voice recognition and machine learning algorithms to provide generalized estimations, in the form of scores extracted from computers, of essential aspects of educator speech. Specifically, the authors assessed the audio quality from teachers' recording with a rating of A, B, C or F. "A" denoted outstanding recording quality, "B" denoted satisfactory quality with minor volume or background noise issues, "C" denoted recordings with flawed segments, and an "F" denoted audio files that were either lost or contained irreparable technical errors. In a comparison with human interpreters, they observed that automatic methods were relatively precise and that voice recognition mistakes had little effect on performance. Therefore, they state that actual instructor conversation can be captured and evaluated for automated feedback. The automated algorithms can also be integrated into/ a dynamic visualization system that will offer educators the necessary feedback for their level of speech. The comparison refers to the evaluation of actual speaker performance as these algorithms utilize voice recognition and machine learning to provide generalized estimations and scores extracted from computers, specifically assessing important aspects of educator speech. This indicates that the focus is on evaluating the content and delivery of the instructor's speech rather than just the technical quality of the audio recording. Furthermore, the feedback offered to educators is tailored to their "level of speech," which suggests that it pertains to assessing how effectively the educators communicate and engage with their students during lectures and conversations in the classroom. In general, these algorithms aim to provide valuable feedback on how well the educators deliver their lectures and interact with their students, helping them improve their communication skills and instructional effectiveness. The approach adopted by Jensen, et al. [23] considers only audio features, and as a result, it does not cater to visual features related to a lecturer's in-class activity.

Jensen et al. [22] also address the issue of designing a framework for automatic educator feedback that necessitates several considerations about audio data harvesting processes, automatic assessment and the way feedback is displayed. The authors employ machine learning techniques, including Random Forest classifiers, alongside transfer learning from the Bidirectional Encoder Representations from Transformers (BERT) algorithm for natural language processing (NLP). BERT, a pre-trained language model based on transformer architecture, is utilized to achieve bidirectional contextual understanding of language, enabling it to capture intricate contextual relationships within sentences. This deep learning model incorporates word embeddings, positional embeddings, and segment embeddings to represent sub word tokens, their positions, and sentence-level distinctions, respectively. The input to both the Random Forest Classifier (RF) and the Bidirectional Encoder Representations from Transformers (BERT) models consists of transcribed utterances. The Random Forest Classifier, features are derived using a bag of n-grams approach, which computes counts of words and phrases (unigrams, bigrams, and trigrams) from the automatically transcribed utterances.

During pre-training, BERT learns contextual representations by solving masked language modeling (MLM) tasks and next sentence prediction (NSP) tasks. The results demonstrate that BERT offers superior and more accurate input across various degrees, rendering it the most practical technique for delivering automated feedback on educator discourse, surpassing the performance of other machine learning techniques, such as Random Forest classifiers.

Zhu et al. [43] propose using the Analytic Hierarchy Process (AHP) and fuzzy decision tree algorithms to evaluate teaching quality, specifically in the context of English education. This system comprises various assessment indexes and procedures to comprehensively evaluate the quality of English instruction. The four critical assessment indexes include Teaching Attitudes, evaluating the disposition and approach of educators; Teaching Contents, focusing on the curriculum's subjects and themes; Teaching Techniques, which inspects the methodologies employed in instruction; and Teaching Effects, analyzing the overall success and efficacy of the educational experience. Utilizing a combination of methods such as student evaluation, peer evaluation, expert evaluation, and self-evaluation, the study integrates these aspects through a decision tree model. The findings contribute to improving teaching quality in colleges by providing objective and reliable evaluation results. However, this study is applicable only to the evaluation of English lessons and cannot be regarded as a generic lecture quality assessment method, as it has not been tested or designed for subjects beyond the scope of English language instruction. Its methodologies and criteria may not translate effectively to other disciplines or linguistic contexts, limiting its broader applicability.

## 3   Defining Lecture Style Quality Indicators

The first step in the proposed methodology for designing a lecture quality evaluation system involves the definition of indicators associated with high and low-quality lecture styles. To define the required features, findings from a thorough research in the existing bibliography in combination with information derived from educators, students and chief education officers were utilized.

As a result of the literature review [4,5,27,39] a set of characteristics related to high and low-quality lecturing styles were gathered. For example, high quality lecturing styles include features related to level of commitment [4], interaction and communication skills [11], promotion of critical thinking, teamwork, and creativity [4], giving directions, helping and giving feedback, avoiding negative words [5], and body gestures [39].

To further refine the set of initial features defined based on the related literature, interviews with stakeholders were staged. During the interviews, participants watched typical YouTube videos showing 'good' and 'bad' examples of lecturing, and they indicated specific lecturer actions regarded as indications of high and low-quality lecturing styles. More specifically participants were shown clips from two online videos with the characteristics of the "*bad* " teacher [1,2] and two online videos with the characteristics of the "*good* " teacher [3,4] . The criteria for selection of the online YouTube videos were based on the search keywords "*good teacher*", "*good lecture* ", "*bad teacher* ", "*bad lecture* " as well as based on rating, number of views and viewer comments.

Interviews were conducted with two chief education officers, two teachers and two students who participated in the video observation action. The views of all three categories of participants are highly important as chief education officers perform the process of educator assessment, while educators and students are directly involved in the educational process. The total duration for each interview was about 30 minutes. After watching each video, participants rated the quality of the lectures they watched, and responded to questions related to the level of lecture quality, the specific actions observed that related to high and low-quality lecturing styles. In addition, participants provided comments on the body language, expression, and speech characteristics of the lectures. More specifically, participants responded the following questions:

**1)** *How would you rate the quality of the lecture and why?*

**2)** *What do you think are the characteristics of a good/bad lecturing style?*

**3)** *Do you think that body language (non-verbal communication) plays an important role in the quality of the lecture and why?*

**4)** *What facial expressions (angry, disgust, fear, happy, neutral, sad, surprise) do you think a good teacher and a bad teacher have during a lecture?*

**5)** *Which actions do you think should be avoided by teachers and which should be used by a good teacher in the classroom?*

**6)** *What do you think are the characteristics of a good teacher in verbal communication and what are the characteristics of a bad teacher in verbal communication?*

Interview data analysis was carried out with the use of codification qualitative techniques [6] in which words, sentences and phrases that had similar conceptual meaning and were important for the research, were arranged in groups whereas data of minor importance were taken apart. This procedure has been repeated to ensure the validity of the codification. Based on the interview findings, most of the participants expressed the view that effective teachers are proficient in using technology, exhibit positive body language (such as facial expressions, head and hand movements), maintain an engaging tone of voice, actively move around the classroom, provide constructive feedback, motivate students, led by example in terms of their attitudes and behavior (such as refraining from using their phone, or drinking in class), are well-prepared, and utilize modern teaching methodologies.

---

[1] https://www.youtube.com/watch?v=7DY1e0Grwtw

[2] https://www.youtube.com/watch?v=3kgtpl4Q5OY

[3] https://www.youtube.com/watch?v=nLfFmYWjHtc

[4] https://www.youtube.com/watch?v=hGBNi4P9OfA

All features defined from the literature and the interviews were assessed with respect to the feasibility of extracting them accurately, and non-invasively, from a video captured using a standard static camera pointing at the lecturer. Furthermore, based on findings from the literature and interview responses, the values corresponding to high and low-quality lecturing types were defined (See Table 2). More information about the features extracted is provided in section 4.

Table 2. Lecture quality metrics.

| Modality | Values for High Quality Lecture Style | Values for Low Quality Lecture Style |
|---|---|---|
| Facial Expressions | Happy, Surprise, Neutral | Anger, Fear, Disgust, Sad |
| Body Activity | Attending, Writing, Hand Raising | Absent, Telephone Call, Texting, Looking Elsewhere |
| Speech | Word Density (35%-55%) Speaking Speed (150-250 words per minute) Speech Intonation (40%-60%) | Word Density (<35%, >55%) Speaking Speed (<150, >250 words per minute) Speech Intonation (<40%, >60%) |
| Hand Movement | Moving | Stationary |
| Facial Pose | Left, Right, Up, Down and Forward | Far-Left, Far-Right, Far-Up, Far-Down, Backwards. |

## 4   Lecture Style Quality Score Estimation

The proposed methodology aims to perform automated recognition of specific measurable characteristics related to verbal and non-verbal communication which are presented in Figure 2. The intensity of the extracted metrics is used for estimating an overall lecture style quality score for each frame and providing feedback to the lecturer. Following the determination of the key features, dedicated techniques were used for extracting those features from video recordings and estimating lecture quality metrics, as shown in Figure 2, and exemplified in the following subsections.

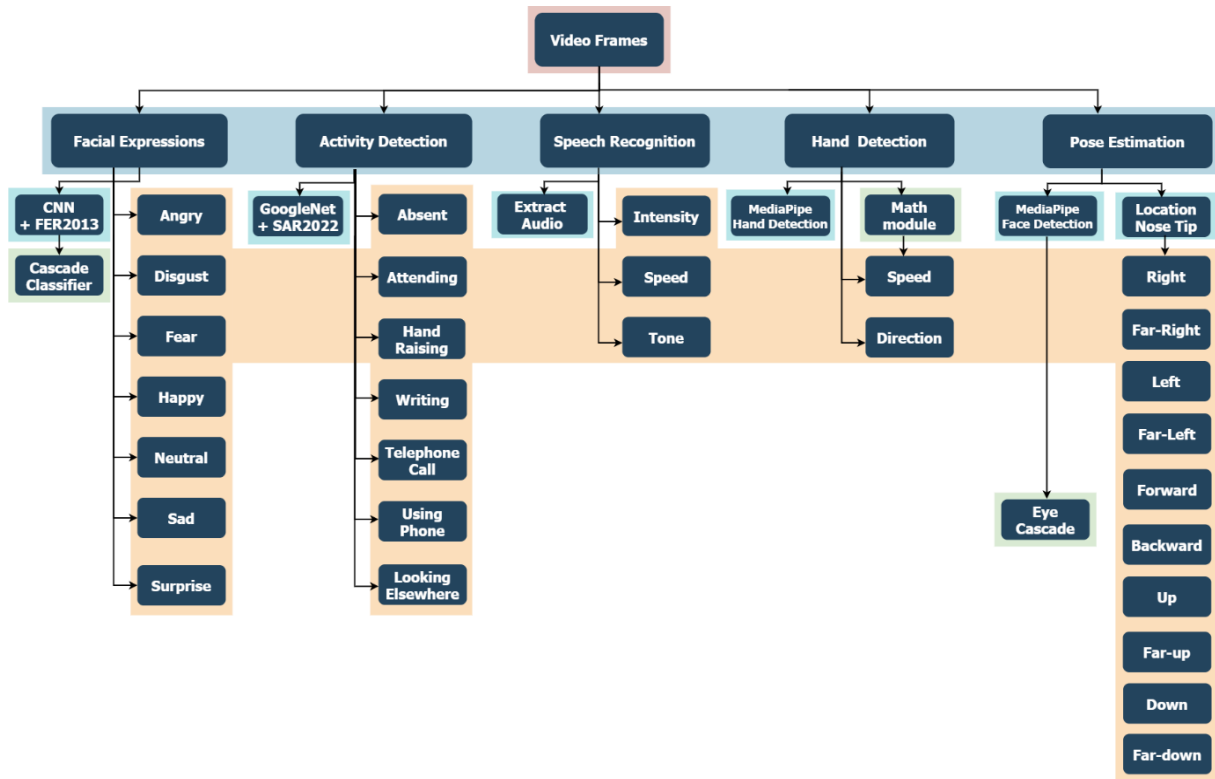

**Figure 2.** Overview of lecture quality features and feature extraction methods.

## 4.1 Facial Expressions

Facial expressions of lecturers are associated with the overall lecture quality, hence in the proposed system facial expression recognition is used. In order to choose the most appropriate method for recognizing facial expressions, a Convolutional Neural Network (CNN) [2], AlexNet [32], ResNet [42] and VGG16 [17] deep networks were trained, and evaluated in recognizing the seven basic emotions of anger, disgust, fear, happy, neutral, sad and surprise. For the training and evaluation, the FER2013 dataset [20] was used. Prior to training, data augmentation in the form of reflection, rotation, scale and translation were used to increase the size of our dataset. Regarding the training hyperparameters, a batch size of 64 was chosen, the learning rate was set at 0.0005 and the Adam optimizer was used. The train set included 28,709 images while the test set included 7,178 images. The results showed that CNN have validation accuracy of 66.45%, which achieved the highest correct recognition accuracy among all networks considered. Figure 3 shows the confusion matrix for expression recognition results using the CNN.

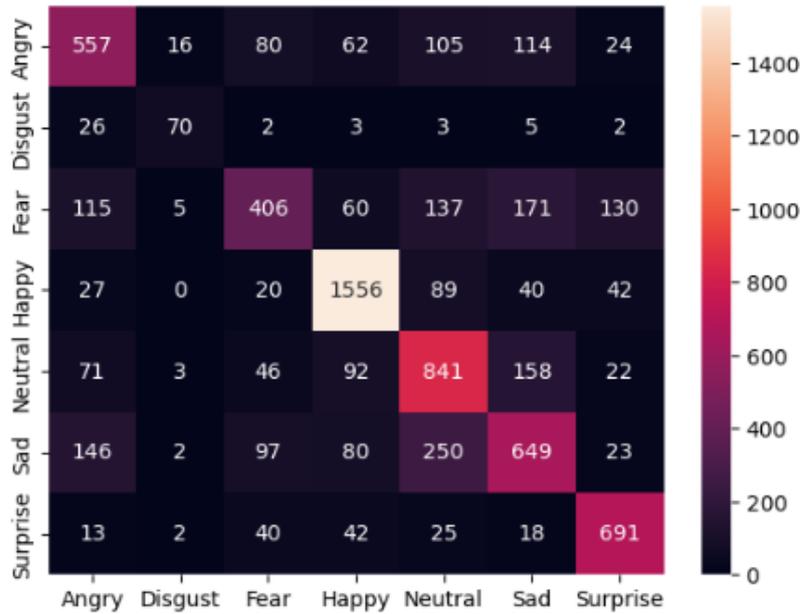

**Figure 3.** CNN confusion matrix.

In the proposed application, expression recognition is performed using a cascade classifier [2] to detect the lecturer's face, and a Convolutional Neural Network (CNN) trained using the FER2013 dataset to classify the seven basics facial expressions (anger, disgust, fear, happy, neutral, sad, surprise) [24]. According to Hou et al. [21] and based on the interviews with the stakeholders, anger, disgust, fear and sad are considered negative emotions while happy, surprise and neutral are considered positive emotions in relation to the teaching process. Therefore, when the expressions of happy, surprise, and neutral are detected a lecture quality score of 1 is assigned, whereas the detection of anger, fear, disgust, and sad expressions leads to a quality score of zero.

## 4.2 Activity Detection

Based on the literature [5,39] and the views of the interviewed stakeholders, the activities performed by a lecturer can contribute towards lecture quality. The proposed system recognizes seven key activities associated with low/high quality lecture styles. In particular the system recognizes the following actions: Absent from the camera point of view, attending, raising hand(s), writing, telephone call, texting, and looking elsewhere. The classification of the seven classes was based on a tuned GoogleNet architecture [14, 16] used for assessing in-class student activity. The proposed system classifies the activities attending, writing and hand raising (see Figure 4) as high-quality lecture styles and assigned a lecture quality score of one, while the activities absent, telephone call, texting, and not looking at the audience (looking elsewhere) are classified as low-quality lecture style (see Table 2), resulting in a lecture quality score of zero.

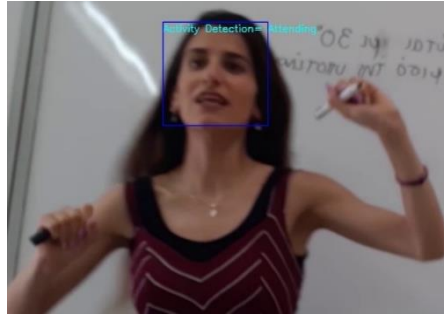

**Figure 4.** Typical screenshot regarding the recognizing teachers' activities in a recorded video.

### 4.3 Speech Recognition

Based on the interviews with the stakeholders, the speech speed, speech intervals and speech intonation are related to lecture speech quality, and for this reason audio characteristics extracted from video segments are utilized in the proposed application. The following features are extracted:

*(a) Word Density*: Non-silent audio intervals in a given speech segment are detected, and the percentage of non-silent against silent parts is estimated providing in that way a metric for Word Density. According to the literature [29], the average word density of teachers is 35-55%, thus word density values between 35%-55% are associated with high quality lecture quality, whereas values outside the recommended range are associated with low quality lecture quality.

*(b) Speaking Speed*: Speaking speed is estimated by dividing the number of words detected in a speech segment over by the segment duration. The range of normal levels of human speech is between 150 and 250 words per minute [25], hence the speaking speed (speed) is correlated with a high-quality lecture style if it is within the normal range, otherwise it is associated with low quality lecture style.

*(c) Speaking Intonation*: Speaking Intonation determines whether the sentence is a question or a statement [19] allowing the derivations of information regarding the level of interaction between a lecturer and the audience. Intonation is estimated as the root mean square (RMS) of the audio waveform. If the mean RMS is greater than 0.01, the intonation is classified as a question; otherwise, it is classified as a statement. The specific value of 0.01 was chosen as a threshold based on experimentation after consulting the librosa library documentation[5]. This approach allows the estimation of the percentage of questions against statements in a given audio segment. A percentage of questions among the 40%-60% in each interval denotes an adequate interaction between the lecturer and the audience hence in that case the quality of lecture is considered high [1], and otherwise it is associated with low quality lecture style.

Based on the three audio-based metrics extracted (Word Density, Speaking Speed, Speaking Intonation), a majority voting approach is employed to determine the overall speech quality. If the majority of the speech-related figures are associated with high-quality lecture style, the corresponding partial lecture speech-based quality score is set to one, whereas in the opposite case the speech-based lecture quality score is set to zero. For the proposed system, the time interval for audio segments was set to three minutes, hence speech metrics related to the intonation, speed and word density are updated every three minutes. An example of the extraction of speech characteristics and estimation of speech quality is shown in Figure 5.

---

[5] https://librosa.org/doc/latest/generated/librosa.feature.rms.html

```
MoviePy - Writing audio in C:/Users/jelen/Speech Recognition/test.wav

MoviePy - Done.
Loading Audio File..
Audio File Loaded..
Calculating..(time depends on file size)
Intensity(Speech Percentage): 96.19%
Speed: 249.00 words per minute
Tone: The recording consists of 94.83% question tone and 5.17% statement tone.
Intensity=  Bad, Speed= Good, Tone= Bad
```

**Figure 5.** Examples of audio characteristics such as speech percentage, speaking speed, and speaking tone estimated from a speech segment.

## 4.4 Hand Movements

Based on the literature hand movements is an important feature of a high-quality lecture [5]. In the proposed system, hand detection is used for estimating the presence of hand movements. Hand detection is performed using Mediapipe Hand Detection [36] and the location of detected hands is used for estimating the speed and direction of hand movements. If a hand is detected in the current frame, a set of landmarks on the hands are localized, and the center of gravity of the landmarks provides an estimate of the hand location. The hand location in each frame is used for estimating the distance and direction of movement of the hand. Hand speed is calculated as the distance between the current center and the previous center of the hand divided by the number of frames elapsed (see Figure 6). Hand direction is calculated as the angle between the current center and the previous center in degrees (see Figure 6). If the hands move in a given interval, hands are classified as "moving" and are considered as a high-quality lecture feature. Otherwise, hands are classified as "stationary", and associated as a low-quality lecture style. The system analyzes hand movements, incrementing the hand movement score by one if hand movement is detected and assigned a score of zero when no movement is detected (see Table 2).

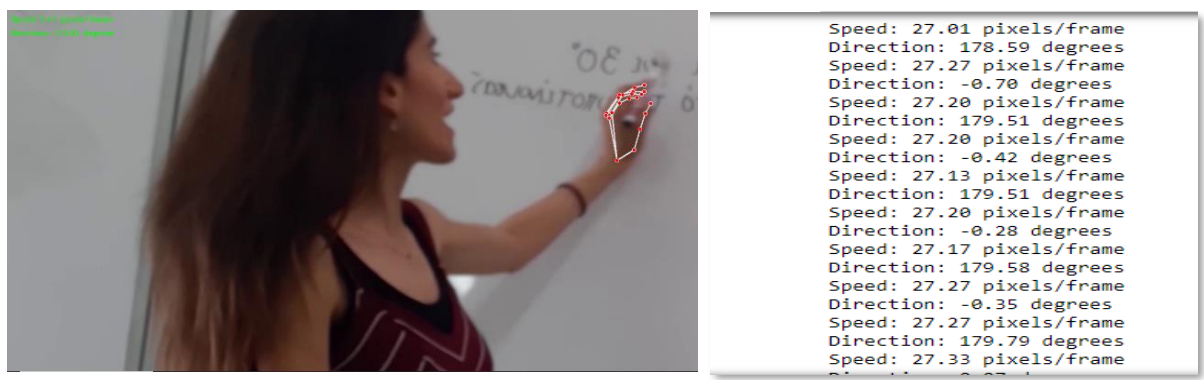

**Figure 6.** Example of hand detection where speed and direction are printed on the console.

## 4.5  Facial Pose Estimation

Facial pose estimation is used for determining if lecturers have optical contact with the camera [37]. Based on the interviews with the stakeholders, facial pose estimation is considered important in education due to its ability to maintain student engagement. Since the camera is positioned from the audience point of view, optical contact with the camera implies optical contact with the audience. Facial pose estimation is performed using Mediapipe Face Detection [34]. The output of the face detector is used for determining if the lecturer is looking *right, left, up, down, far-left, far-right, far-up, far-down, forward and backwards.* If a face is detected in the current frame, the nose tip is also located. The position of the nose tip relative to the bounding box around the face allows the calculation of the direction that the lecturer is looking (see Figure 7). In addition, an eyeCascade detector was used to detect eyes in a video frame. If the algorithm does not detect eyes, it assumes that the person is not looking towards the audience. Bearing in mind that the camera is positioned in front of the lecturer, when the lecturer is looking right, left, up, down and forward are classified as facial poses associated with high quality lecture style, as in that case it is implied that there is eye-contact between the lecturer and the audience. When far-right, far-left , far-up, far-down and backwards are detected the head direction is associated with a low-quality lecture style, as this implies that there is no optical contact with the audience (see Table 2).

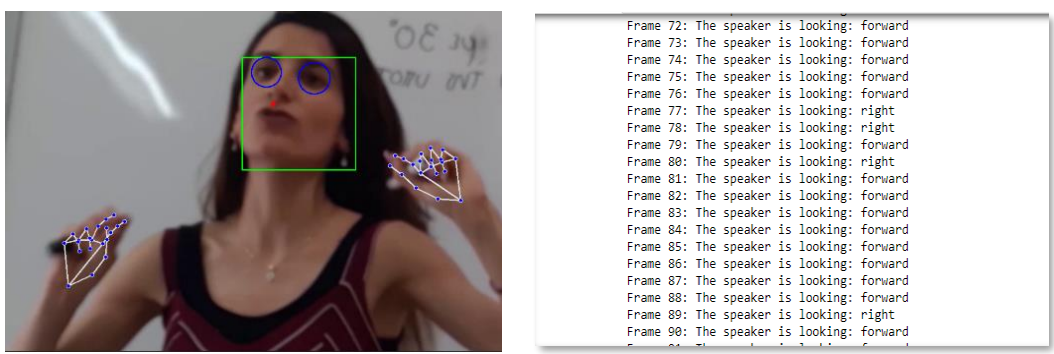

**Figure 7.** Example of pose estimation where direction is printed on the console.

## 4.6  Merging Metrics

Overall, the proposed system utilizes five modalities (Facial Expressions, Body Activity, Speech, Hand Movement, and Facial Pose), and a positive measure for a given modality increases the total quality score by one. As a result, possible values for the quality score range between zero to five, where zero means that none of the modalities resulted in an acceptable score regarding lecture quality and a score of five indicates that all modalities produced a score associated with a high-quality lecture. The total score is calculated by summing the scores of each frame and dividing by the number of frames, and the total score is presented on the console providing real-time feedback to the lecturer (see Figure 9(a)). The process of estimating the overall quality score from partial scores is illustrated in Figure 8. Figure 9, shows an example of the overall lecture quality graph, activity score, facial expressions score, facial pose score, hand movement score, score per frame and speech score for a given part of a lecture.

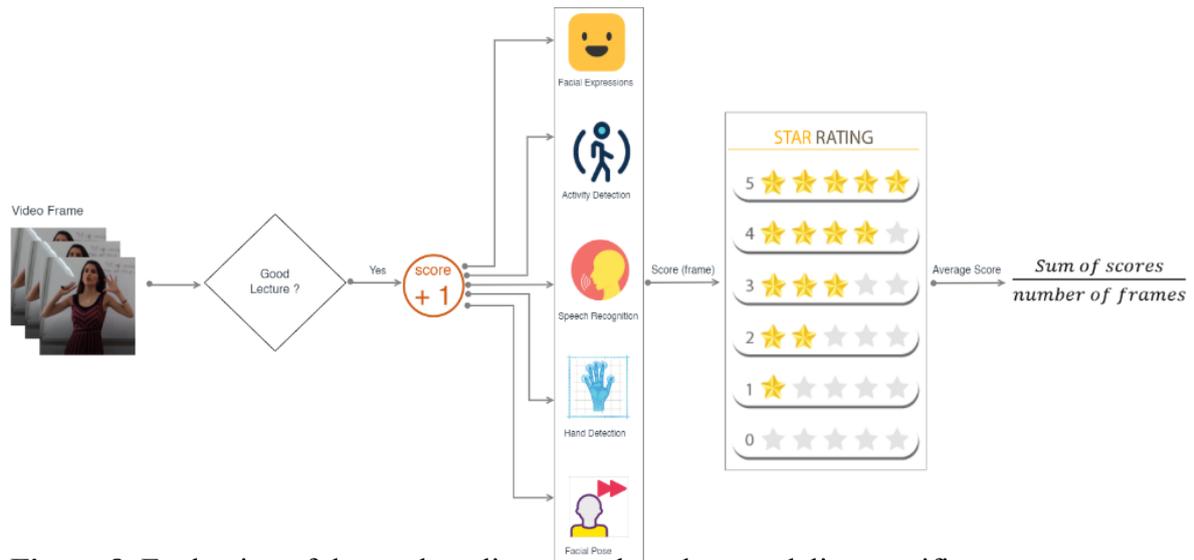

**Figure 8.** Evaluation of the total quality score, based on modality-specific scores.

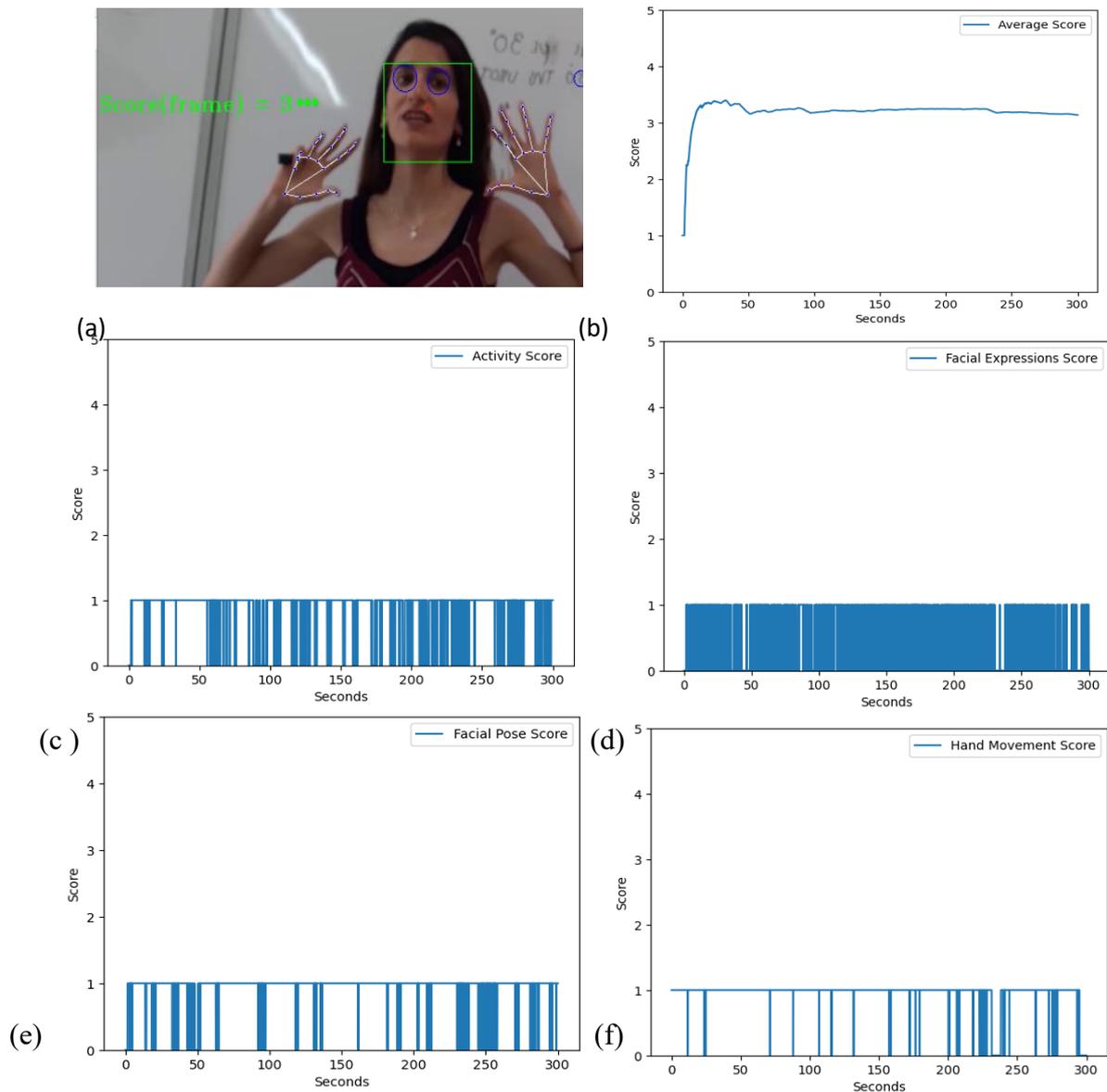

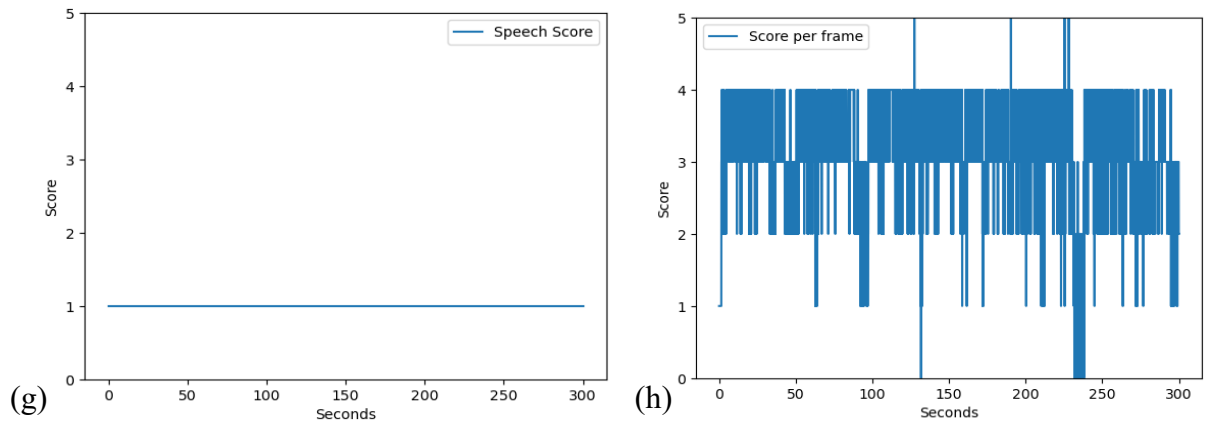

(g)  (h)

**Figure 9.** (a) Score for each frame, (b) Average score, (c) Activity score, (d) Facial Expressions score, (e) Facial Pose score, (f) Hand Movement score, (g) Speech Score printed on the console, and (h) Score per frame.

Indicative frames classified as high- or low-quality lecture style are shown in Figure 10. The image frame in Figure 10(a) is considered to show a high-quality lecture-style because it includes positive facial expressions (happy), positive activity detection (hand raising), hand movements, and facial pose (forward) that ensures eye-contact with the audience. For that interval speech features were not among the range of acceptable values, hence the overall score was four out of five. The image frame in Figure 10(b) was assigned a low lecture-style quality score because no positive facial expressions were detected, a negative action (looking elsewhere) and facial pose (backward) were detected, and the speech features were not among the range of acceptable values. However, since hand movements were detected, the score was assigned a score of one out of five. The image frame in Figure 10(c) was classified to a low-quality lecture-style quality style because a negative activity was detected (making a telephone call), while and facial pose (far-right), and the speech features were not among the range of acceptable values. However, since hand movements and a positive facial expression (neutral) was detected, the overall score was estimated to a value of two out of five.

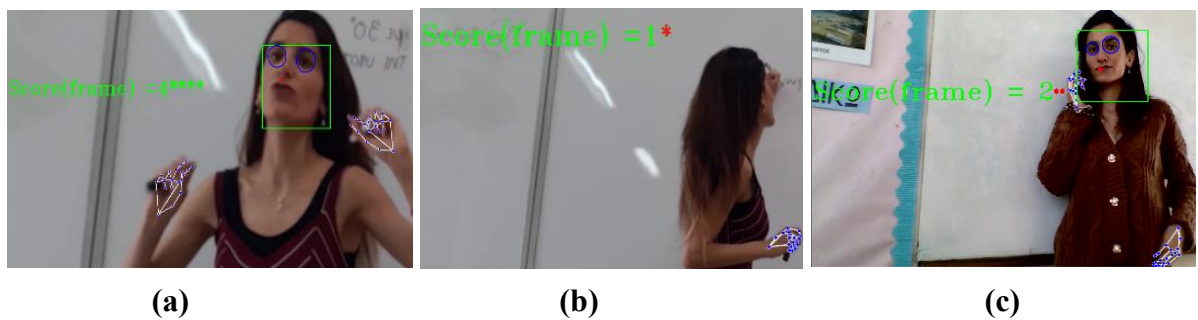

(a)  (b)  (c)

**Figure 10.** Typical screenshots where high (a) or low-quality lecture styles (b)-(c) are detected.

The proposed system provides two options for providing feedback to the lecturer. The first option estimates the quantifiable metrics in real-time using the laptop camera to capture the video, so that low quality scores can be used for alerting the attention of the lecturer. The second option loads a recorded video of a lecture and produces quality score estimations for the overall lecture so that the lecturer can study and reflect about his/her performance at the end of a lecture.

The overall application operates efficiently in cases where the camera is located at a close distance from the speaker and the microphone is close to the speaker. In cases, where the speaker is at a far distance from the camera/microphone, proper detection of the body parts and extraction of audio features cannot be achieved accurately. Hence, in its current form the proposed system applies to cases where the lecturer does not move in a class, and the camera is located at a reasonable distance capturing images of the lecturer from the audience point of view. The use of suitable microphones attached on the lecturer is recommended.

## 5 Performance Evaluation

In this section a performance evaluation of the proposed lecture style quality evaluation system is presented. The evaluation focuses on examining the usefulness of the proposed system, determining whether the lecture quality values generated by the proposed system are accurate and aligned with the views of human evaluators.

## 5.1    Preliminary Interview based-evaluation

To investigate the acceptance of the proposed methodology, a preliminary user evaluation using semi-structured interviews has been carried out with two chief education officers, two teachers and two students. The interviews were conducted with the same participants who participated in the user requirements analysis step. One of the main tasks of a chief education officers relates to the guidance and evaluation of teaching staff in schools. Consequently, their opinion plays a crucial role in developing and completing the application.

Initially, the researchers demonstrated the application to the participants in real-time in order to receive more reliable feedback. Subsequently, participants answered questions regarding their impressions, about biometric features, and possible ways to improve the application. More specifically, participants responded the following questions:

1) *What are your impressions about the application*?
2)  *Is the application considered useful, and if so, why*?
3)  *Can you suggest possible ways to improve the application*?

The results showed that all participants consider the system to be an innovative idea as this system can provide real-time feedback. Participants also stated that features used accurately reflect the lecture style quality. Regarding ways to improve, participants mentioned that the camera may not only focus on the teacher but also on the students so that student feedback is also considered as part of the evaluation process. Based on the positive outcome of the preliminary interview-based evaluation that ensures the usefulness of the proposed system, we proceeded to the process of quantitative evaluation of the performance of the system, as described in the following sections.

## 5.2 Quantitative Evaluation

In order to stage a quantitative evaluation of the system, ground truth data is required. Due to the lack of such data, an annotation process was staged with nine annotators aged from 18 to 62 years. Annotators included university students who study in public and private institutions and educators of elementary, secondary and tertiary levels. The average time needed for each annotator to complete the annotation process was approximately 60 minutes. More details about the process are provided below.

### 5.2.1 Data Annotation

The participants were asked to complete a three-stage annotation process. In the beginning, the participants were asked to carefully read the agreement form in order to give their consent regarding their answers which will remain confidential and will be exclusively used for the purposes of this research. The first part of the of the annotation process included a series of questions related to the annotators (ethnicity, gender, age, occupation) as well as open type questions, where annotators provided information about their educational experience (if any), field of expertise, while students were called to define the field, the level and the duration of their studies.

At the beginning of the other two parts (second and third) annotators were given instructions related to the five biometric features extracted by the system (i.e. facial expressions, activity performed, head direction, hand movements, speech recognition), in order to make their evaluation clearer and simpler. During the second part of the annotation procedure, annotators were presented with 100 image frames extracted from nine different lecture videos showing lectures with varying quality of presentation. Frames used were selected to portray a variety of different conditions reflecting different teaching styles. Each annotator had to evaluate the perceived lecture quality in each frame with respect to the following biometric features: Facial expression, educators' actions, hand movements, and head movements. Annotators also had to provide an overall evaluation of the lecture quality in each frame using the 4 level Likert scale, ranging from "very low" to "very high" lecture style quality.

The third part of the online questionnaire contained 100 sound segments with a duration of 15 seconds each, which were extracted from the nine videos used in part B of the evaluation. In this part, the participants listened each speech segment and provided an estimate of speech quality for each segment in the 4 level Likert scale ranging from "very low" to "very high" lecture style quality.

For the evaluation of the application, the data annotated was used as the ground truth. In this study, the average among all nine annotators was considered as the ground truth. The four-class problem for the facial expressions, activity detection, speech recognition and facial pose was transformed to a binary problem by considering annotations values of 1 and 2 as low-quality lecture, and the values 3 and 4 as high-quality lectures. The prevailing mode for every biometric feature and the total score per frame were calculated. The prevailing mode considered as the ground truth for each biometric feature is defined as the one that appears most often on the dataset.

### 5.2.2 Results

In this section, the performance evaluation results of our application and the results for assessing human performance are presented. Subsequently, machine and human results are compared in order to obtain a more complete picture about the performance of the proposed application.

#### 5.2.2.1 Performance Evaluation Results

For each visual/speech sample used, the output values given by the proposed system in every frame and the total score for every frame were calculated, and compared with the ground truth obtained from the annotators (see section 5.2.1). To evaluate the performance of the proposed model, specific performance measures for each biometric feature such as Accuracy, Precision, Recall, MCC, Cohen's kappa and F1-score were calculated. F1-score is a combination of the Precision (true positives / total predicted positives) and the Recall value (true positives/total actual positives). Furthermore, the confusion matrix for every metric is shown in Figure 11.

Table 3: Accuracy metrics regarding the comparison between actual and output values.

| Performance Measures | Expressions | Body Activity | Facial Pose | Hand Movements | Speech | Score (frame) |
|---|---|---|---|---|---|---|
| Accuracy(%) | 83.0 | 73.0 | 70.0 | 75.0 | 69.0 | 72.0 |
| Recall | 0.830 | 0.730 | 0.700 | 0.750 | 0.690 | 0.720 |
| Precision | 0.829 | 0.784 | 0.688 | 0.746 | 0.711 | 0.714 |
| F1-score | 0.821 | 0.729 | 0.689 | 0.741 | 0.684 | 0.706 |
| MCC* | 0.593 | 0.515 | 0.309 | 0.445 | 0.402 | 0.652 |
| Cohen's k | 0.578 | 0.479 | 0.304 | 0.435 | 0.383 | 0.648 |
| Error | 0.17 | 0.27 | 0.30 | 0.25 | 0.31 | 0.28 |
| **Mean Absolute Error (MAE)** | N/A | N/A | N/A | N/A | N/A | **0.44** |

*MCC=Matthews Correlation Coefficient*

Based on the results it is evident that the algorithm can classify teachers' body language with an accuracy of 72% per frame. From the confusion matrices, the proposed system recognized the facial expressions with 83% accuracy, while the other metrics have accuracy higher than 69%. Figure 11 represents the confusion matrix for each biometric feature and score (per frame). The prediction error is relatively small therefore the performance satisfies the requirements for a practical application [40].

Regarding the facial expression confusion matrix (see Figure 11), t system mostly misclassifies the class "High" as " Low" quality with percentage of 4% and the class "Low" as "High" quality with percentage of 13%. Similarly, the proposed system mostly misclassifies the class "High" as "Low" quality with percentage of 4% and the class "Low" as "High" quality with percentage of 23% regarding the body activity. From the confusion matrices of facial pose and hand movements (see Figure 11), the proposed system mostly misclassifies the class "High " as "Low " quality with percentage of 11% and 8% respectively. Otherwise, the proposed system misclassifies the class "Low" as "High" quality with percentage of 19% and 17% respectively.

Regarding the speech recognition, the system misclassifies the class "High" as "Low" quality with percentage of 8% and the class "Low" as "High" quality with percentage of 23% (see Figure 11). In conclusion, the proposed system is doing better in recognizing expressions associated with high quality lecture styles rather than expressions associated with low quality lecture styles.

From the confusion matrix in Figure 11, the proposed system perfectly classifies frames with score "Zero". Our system mostly misclassifies the score "One" as "Two" with percentage of 2% and the score "Two" as "Zero", "One", Three" with percentages of 2%,3%,3% respectively. Furthermore, the system misclassifies the score "Three" as "Zero", "One", "Two" with percentages of 3%,4%,5% respectively. The score "Four" misclassifies as the score "One", "Two", "Three" with percentages of 1%, 1%, 3% respectively and the class "Five" misclassify as the class "Three" with percentage of 1%. The scores "Zero," "One," "Two," and "Three" are considered to represent low-quality lecture styles, while the scores "Four" and "Five" are considered to represent high-quality lecture styles. Hence, we observe that the percentages that misclassify frames and audio are less than 5% and thus our proposed system rarely misclassifies frames and audio. The score per frame is 72%, which means that in 72 out of 100 cases, the system correctly classifies the teaching quality. Furthermore, regarding the score per frame, the Mean Absolute Error (MAE) calculated as the mean of the sum of the absolute difference between the actual values and predicted values. The average difference between the actual values and predicted values by the model is 0.44 regarding the score per frame. This indicates that the predicted values will be off by an average of 0.44 units for measurements in the range of 1 to 4.

(a)  (b)

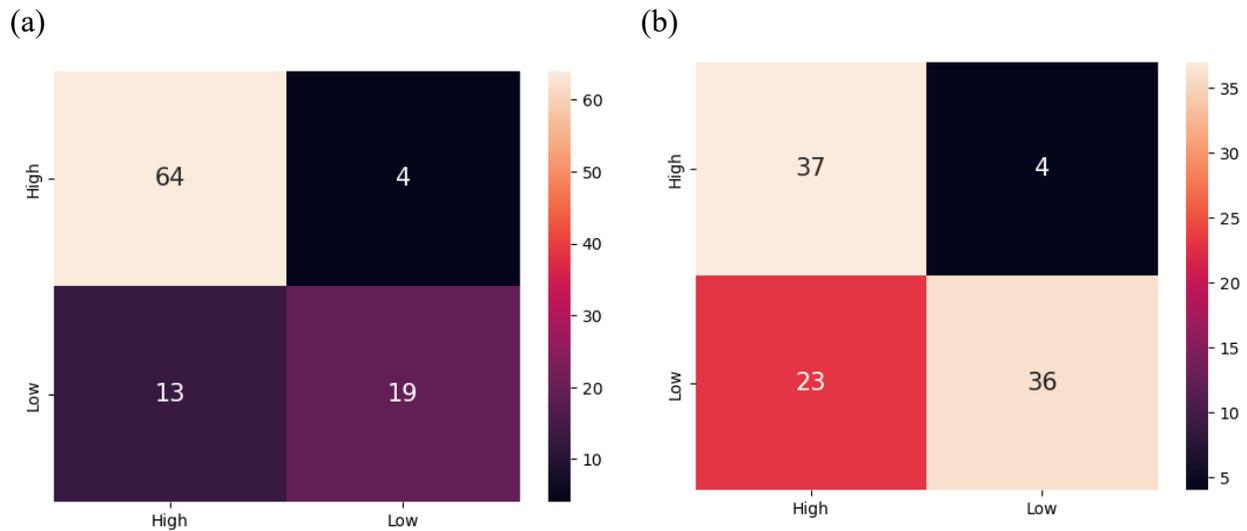

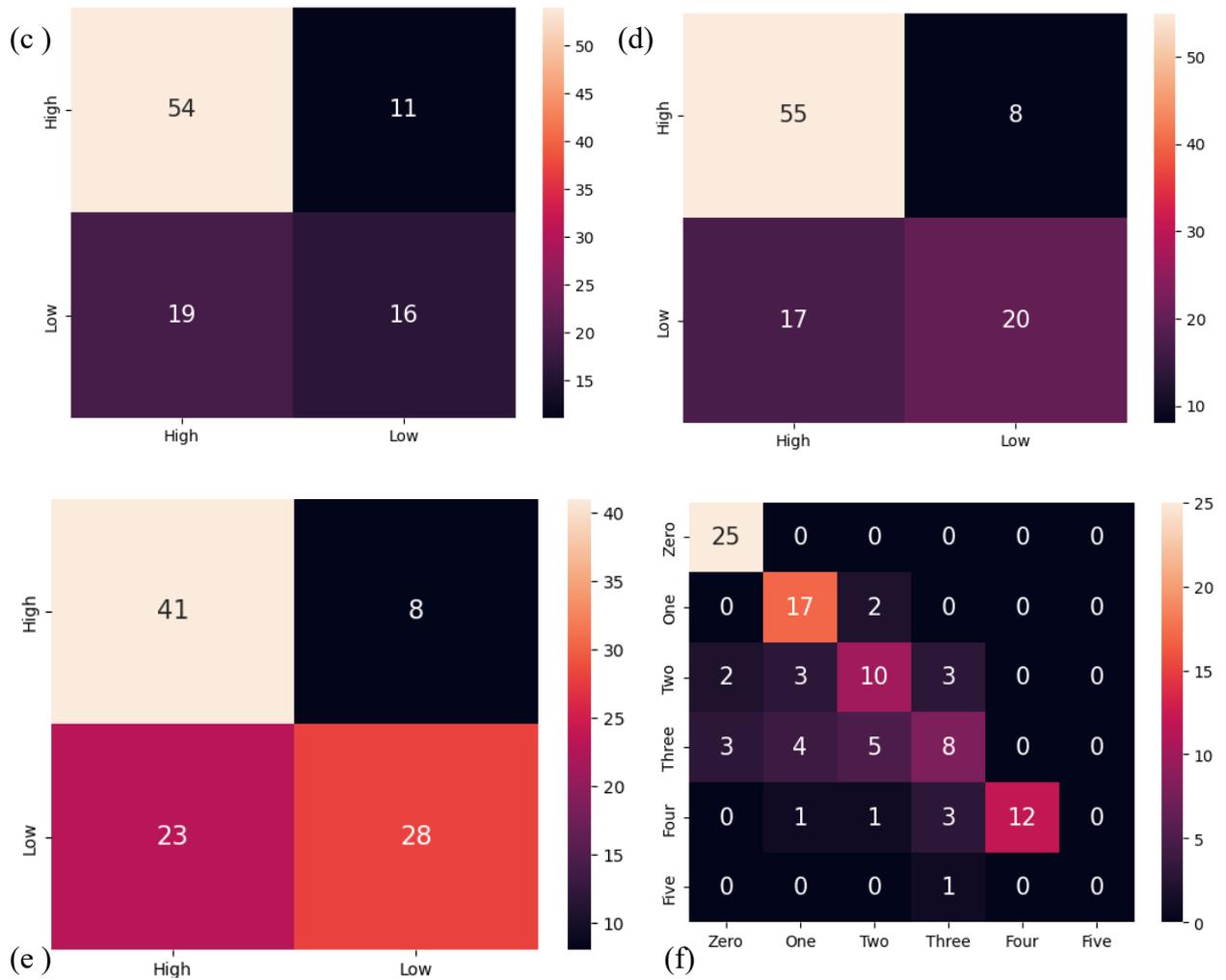

**Figure 11.** Confusion matrix for (a) facial expressions (first row-left), (b) body activity (first row-right), (c) facial pose (second row-left), (d) hand movements (second row-right), (e) speech (third row-left) and (f) score per frame (third row-right).

### *5.2.2.2 Assessing human performance*

The task of lecture quality assessment is a subjective task, hence before concrete conclusions related to the performances of the proposed system are derived, it is essential to derive conclusions related to the level of conformity among the human annotators. For this purpose, leave-one-out experiments were carried out among the nine annotators where in turn the responses of the eight participants were considered as the ground truth and tested against the performance of the remaining annotator. The Mean Absolute Error (MAE) obtained among the nine leave-one out tests was 0.602. This indicates that the difference in responses among the participants will be off by an average of 0.602 units in the 1 to 4 scale. Each participant has their own opinion, so it is reasonable to expect differences in their responses for the subjective task of lecture style quality evaluation. The overall accuracy metrics were calculated by comparing the ground truth values to the values of the nine participants for each type of features and the overall quality score per frame (see Table 4).

Table 4: Overall accuracy metrics for the nine annotators.

| Performance Measures | Expressions | Body Activity | Facial Pose | Hand Movements | Speech | Score (frame) |
|---|---|---|---|---|---|---|
| Accuracy(%) | 83.6 | 77.6 | 81.3 | 75.8 | 66.0 | 67.3 |
| SD* (Accuracy) | 0.07 | 0.07 | 0.12 | 0.09 | 0.02 | 0.07 |
| Recall | 0.836 | 0.776 | 0.813 | 0.758 | 0.660 | 0.673 |
| Precision | 0.837 | 0.793 | 0.819 | 0.766 | 0.670 | 0.725 |
| F1-score | 0.836 | 0.777 | 0.815 | 0.760 | 0.653 | 0.687 |
| MCC* | 0.626 | 0.561 | 0.600 | 0.496 | 0.327 | 0.602 |
| Cohen's k | 0.626 | 0.552 | 0.599 | 0.494 | 0.316 | 0.598 |
| Error | 0.16 | 0.22 | 0.19 | 0.24 | 0.34 | 0.33 |
| **Mean Absolute Error (MAE)** | N/A | N/A | N/A | N/A | N/A | **0.60** |

*SD=Standard Deviations, MCC=Matthews Correlation Coefficient,

Based on the results it is evident that the humans can classify teachers' body language with accuracy 67.3% per frame. From the confusion matrices, the humans recognized the facial expressions with 83.6% accuracy, while the other metrics have accuracy higher than 66%. The standard deviation of the nine annotators' accuracy was calculated, where the values were 0.07, 0.07, 0.12, 0.09, 0.02, and 0.07 for facial expressions, activity detection, speech recognition, hand movements and facial pose respectively. The standard deviation of accuracy describes how much the accuracy values of the population differ from the mean accuracy value. It is observed that there is a variance between people in terms of classification accuracy in each metric due to the fact that each participant has their own opinion. Figure 12 represents the confusion matrix for each biometric feature and score (per frame).

Regarding the facial expression confusion matrix (see Figure 12), humans mostly misclassify the class "High" as "Low" quality with percentage of 8.8% and the class "Low" as "High" quality with percentage of 7.7%. Similarly, humans mostly misclassify the class "High" as "Low" quality with percentage of 6.6% and the class "Low" as "High" quality with percentage of 15.9% regarding the body activity. From the confusion matrices of facial pose and hand movements (see Figure 12), humans mostly misclassify the class "High" as "Low" quality with percentage of 11% and 14.4% respectively. Otherwise, the proposed system misclassifies the class "Low" as "High" quality with percentage of 7.7% and 9.8% respectively. Regarding speech recognition, humans misclassify the class "High" as "Low" quality with percentage of 23.4% and the class "Low" as "High" quality with percentage of 10.6% (see Figure 12). In conclusion, expressions associated with low quality lecture styles rather than expressions associated with high quality lecture styles are recognized by humans with high accuracy. From the confusion matrix in Figure 12, humans mostly misclassify the score "Zero" as "One", "Two", "Three", "Four", "Five" with percentage of 2%, 3.3%, 2.2%, 1.2%, 0.7% respectively and the score "One" as "Zero", "Two", "Three", "Four", "Five" with percentage of 0.2%, 1.6%, 1.6%, 0.4% and 0.6% respectively. Furthermore, humans misclassify the score "Two" as "One", "Three", "Four", "Five" with percentage of 1.4%, 2.3%, 1.4%, 0.2% respectively. The score "Three" misclassifies as the score "Zero", "One", "Two", "Four", "Five" with percentages of 0.3%, 0.9%, 2.1%, 1.8% and 1.3% respectively and the class "Four" misclassify as the class "One", "Two", "Three", "Five" with percentage of 0.3%, 0.9%, 3.9% and 1.4% respectively. The score "Five" misclassifies as the score "Two", "Three", "Four", "Five" with percentage of 0.1%, 0.1%, 0.2% and 0.6% respectively. Hence, we observe that the misclassification percentages are less than 4% and thus our proposed system rarely misclassifies frames and audio.

Regarding the score per frame, the Mean Absolute Error (MAE) calculated as the mean of the sum of the absolute difference between the actual values and predicted values. The average difference between the actual values and predicted values by the model is 0.60. The accuracy score regarding the score per frame is 67%, which means that in 67 out of 100 cases, the humans correctly classify the teaching quality.

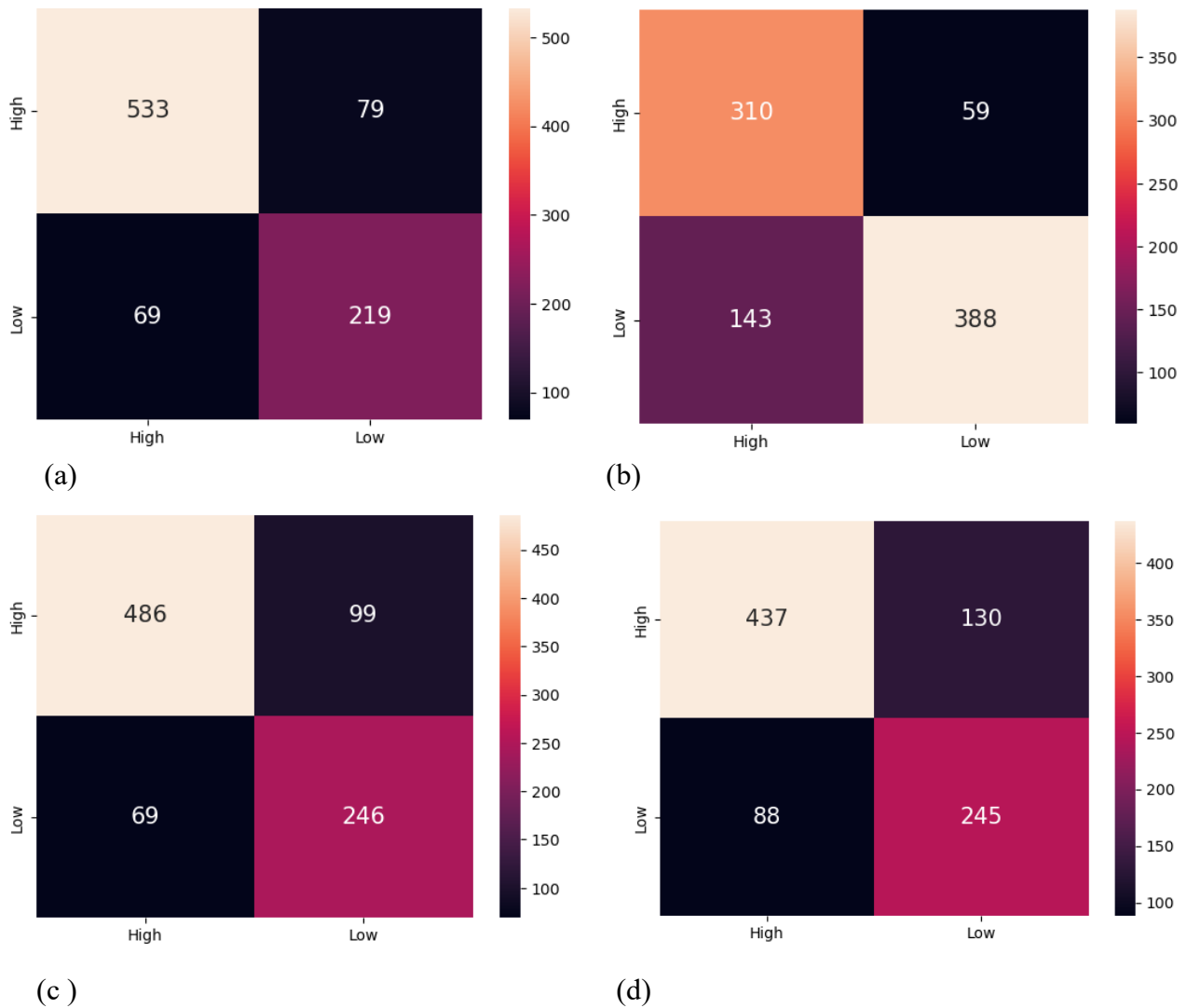

(a)

(b)

(c )

(d)

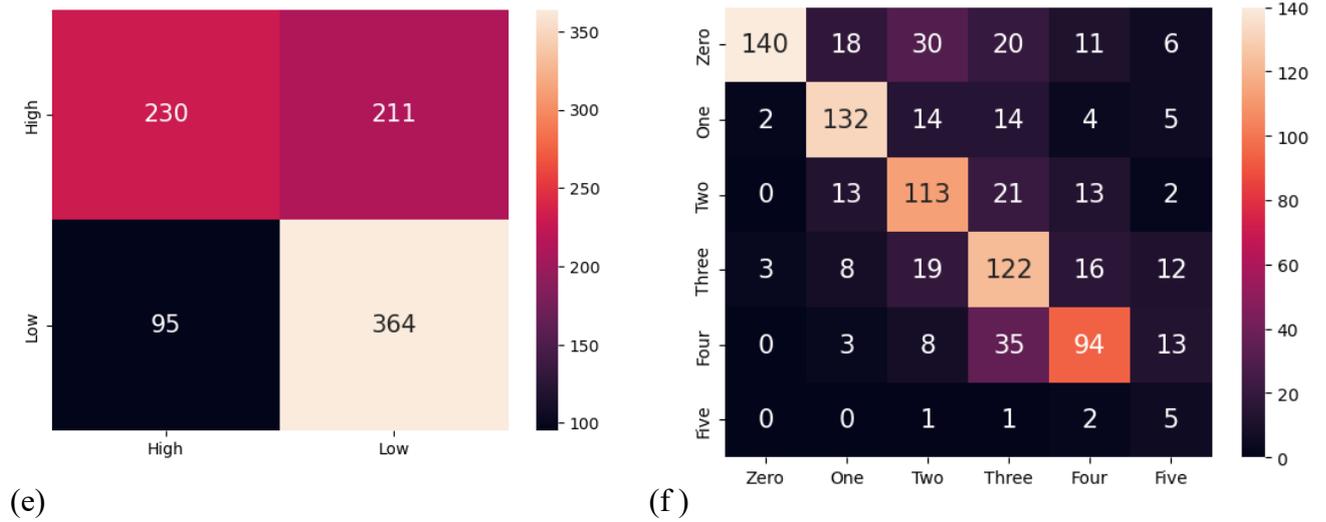

(e)                                    (f)

**Figure 12.** Confusion matrix for (a) facial expressions (first row-left), (b) body activity (first row-right), (c) facial pose (second row-left), (d) hand movements (second row-right), (e) speech (third row-left) and (f)score per frame (third row-right).

### *5.2.2.3 Comparison machine and human performance*

In this subsection, the goal is to compare the performance of the proposed system with human performance. In order to investigate differences between the mean accuracy of people and machine in variables "Expressions", "Body Activity", "Facial Pose", "Hand Movements", "Speech" and "Score per frame" the following hypotheses were examined.

**Ho :** Human performance is equal to the machine performance in the task of estimating lecture quality style.

**H₁ :** Human performance is different from machine performance in the task of estimating lecture quality style.

Inferential statistics were applied for exploring the differences between the groups and for testing the hypothesis using the Chi-square test, as the samples are independent, and the data are in ordinal and nominal scales. Subsequently, the Holm-Bonferroni method was used in order to deal with the problem of multiple comparisons. The significance level was set at 5%. The null hypothesis is accepted when p-value≥0,05 and is rejected when p-value<0.05.

Subsequently, the Holm-Bonferroni adjusted Chi-square test was used to compare the mean absolute errors for each modality. The results show that there is no statistically significant difference regarding the facial expressions (p=1.00), facial pose (p=0.696), hand movements (p=0.696), speech recognition (p=0.161) and overall score per frame (p= 1.00) (see Table 5). Therefore, human performance is equal to the machine performance for teachers' body language assessment regarding the facial expressions, facial pose, hand movements, speech recognition and overall score per frame. Hence, the overall score per frame of the machine are the same with the overall score per frame of the human. On the other hand, there was a statistically significant difference regarding the biometric feature activity detection (p= 0.042<0.05). In Table 5, we can observe that the human's mean absolute error regarding the activity detection is 0.59 while the machine's is 0.40.

Table 5. Mean absolute error, standard deviations for each group, and comparison of groups for each biometric feature and overall score per frame.

| Biometric Features | Humans | Machine | Comparison |
|---|---|---|---|
|  | M (SD) | M (SD) | p-value |
| Facial Expressions | 0.32 (0.469) | 0.35 (0.479) | 1.000 |
| Activity Detection | 0.59 (0.494) | 0.40(0.492) | **0.042*** |
| Speech Recognition | 0.51(0.502) | 0.36(0.487) | 0.161 |
| Hand Movements | 0.37(0.485) | 0.28(0.451) | 0.696 |
| Facial Pose | 0.35(0.479) | 0.27(0.446) | 0.696 |
| Score per frame | 1.88(1.472) | 1.54(1.374) | 1.000 |

SD, standard deviation
**\*Statistically significant change (p value<0,05).**

Therefore, we conclude that the performance of the proposed system is similar to the performance of human annotators for lecture style quality estimation based on facial expressions, facial pose, hand movements, speech recognition and the overall quality score per frame. The only case that the performance between the machine and annotators was statistically different, was in the case of activity detection. However, even in that case the mean absolute error of the proposed application (0.44) (see Table 3) is lower and the mean absolute error for humans (0.6) (see Table 4), hence in the task of activity detection there is evidence that the proposed system outperforms humans. In conclusion, based on the results of statistical tests, the performance achieved by the proposed system is either similar or even better to the performance of human annotators.

## 6   Conclusions and Future Work

A pilot automated system for evaluating the quality of the educators' lectures, based on the definition and extraction of a set of biometric features was presented. The proposed application extracts multiple features such as facial expressions, body activity, hand movement, facial pose, speech word density, speaking speed, and speech intonation, which are combined to provide a lecture style quality score for each frame as well as an overall score for the whole lecture.  The acceptance of the application was evaluated by chief education officers, teachers and students regarding the functionality, usefulness of the application, and possible improvements. The results showed that participants found the application novel and useful in providing automated feedback regarding lecture quality, so that the overall teaching process is benefitted.

In addition to the interview-based user evaluation, a quantitative performance evaluation was staged, where the accuracy of estimating lecture style quality using different modalities was evaluated. As part of the process, the performance of the system was tested against the ground truth obtained based on the opinions of several annotators. The performance of the system proved the feasibility of using it in real applications. Furthermore, the performance of the system was compared to the performance achieved by humans in the task of lecture style quality estimation. The results from the machine-human comparison showed that our proposed system achieves a similar performance to  humans regarding the modalities of  facial expressions, facial pose, hand movements, speech recognition and score per frame. Furthermore, human and machine performance analysis showed that the performance of our proposed system outperforms the performance of humans regarding the speech (69% and 66% respectively) and score per frame (72% and 67.3% respectively). The performance of our system differs little from that of humans regarding the biometric feature body activity. This is due to the fact that humans' performance evaluation was based on everyone's personal opinion, therefore, it is considered reasonable that there is some variation in the evaluation of metrics between humans and the machine. Nevertheless, the proposed system achieves smaller mean absolute error in estimating lecture style quality features associated with a teacher's body language.

In the future, we plan to further refine the quality metrics considered, and their combination to produce the total quality score, while the application will be thoroughly re-evaluated. Furthermore, based on the user evaluation comments, additional cameras, and cameras attached on the lecturer, will be used so that the cases that the lecturer moves in a class are considered. In the future, we also plan to conduct evaluation experiments during real-time courses in order to examine the improvement of teachers lecture style and possible improvement in class performance resulting from the use of the proposed application, so that the pedagogical added value of the application in real conditions is cross-verified.